\documentclass[letterpaper, 11 pt, onecolumn, conference]{cmds}  

\IEEEoverridecommandlockouts     
\usepackage{latexsym}		
\usepackage{graphicx}		
\usepackage{fixltx2e}   
\usepackage{rotating}		
\usepackage{amsfonts}%
\usepackage{epsfig,epstopdf} 
\usepackage{amsmath,amssymb} 
\usepackage{float} 
\usepackage{tcolorbox}
\usepackage[hidelinks]{hyperref} 
\usepackage{color}
\newtheorem{theorem}{Theorem}

\newtheorem{remark}{Remark}
\newtheorem{fact}{Design Properties}

\newtheorem{definition}{Definition}

\newtheorem{design}{Design Procedure}

\usepackage[ruled,vlined]{algorithm2e}

\usepackage{tikz}
\usetikzlibrary{calc,patterns,decorations.pathmorphing,decorations.markings}
\usepackage{tcolorbox}
\usepackage{preview}
\newcommand{\yslant}{0.5}
\newcommand{\xslant}{-0.6}
\definecolor{darkgreen}{rgb}{0.1,0.8,0.5}
\definecolor{darkbrown}{rgb}{1.0, 0.25, 0.25}
\usetikzlibrary{positioning}
\usetikzlibrary{arrows}
\usepackage{subcaption}
\newsavebox{\tempbox}

\usepackage{ mathrsfs } 

\usepackage[english]{babel} 
\title{\LARGE \bf Arbitrarily Fast Switched Distributed Stabilization of Partially Unknown Interconnected Multiagent Systems:\\ A Proactive Cyber Defense Perspective 
}

\author{Vahid Rezaei$^\star$, Jafar Haadi Jafarian$^\star$, and Douglas C. Sicker$^{\star \dagger}$
\thanks{\noindent $^{\star}$ V. Rezaei, \textcolor{black}{J. H. Jafarian}, and D. Sicker are with the Department of Computer Science and Engineering, University of Colorado, Denver, CO 80204, USA, Emails: {\tt \small firstname.lastname@ucdenver.edu}.}
\thanks{\noindent $^{\dagger}$ D. Sicker is with the Department of Electrical Engineering, University of Colorado, Denver, CO, USA.}
}
\begin{document}

\maketitle
\thispagestyle{plain}
\pagestyle{plain}

\begin{abstract} 
	A design framework recently has been developed to stabilize interconnected multiagent systems in a distributed manner, and systematically capture the architectural aspect of cyber-physical systems. Such a control theoretic framework, however, results in a stabilization protocol which is passive with respect to the cyber attacks and conservative regarding the guaranteed level of resiliency. We treat the control layer topology and stabilization gains as the degrees of freedom, and develop a mixed control and cybersecurity design framework to address the above concerns. From a control perspective, despite the agent layer modeling uncertainties and perturbations, we propose a new step-by-step procedure to design a set of control sublayers for an arbitrarily fast switching of the control layer topology. From a proactive cyber defense perspective, we propose a satisfiability modulo theory formulation to obtain a set of control sublayer structures with security considerations, and offer a frequent and fast mutation of these sublayers such that the control layer topology will remain unpredictable for the adversaries. We prove the robust input-to-state stability of the two-layer interconnected multiagent system, and validate the proposed ideas in simulation.
\end{abstract}
\section{Introduction}\label{Section:Introduction}
In response to the advances in embedded sensing, computation, and wireless communication, multiagent and cyber-physical systems (MASs and CPSs) have attracted significant attention during the past two decades. The preliminary studies in the literature of MASs were mainly focused on the simple integrator or linear time-invariant (LTI) agent models in order to achieve consensus in MASs, by creating a meaningful connection between control and graph theories~\cite{KnornChenMiddleton-TCNS-2015}. Later, this attention was shifted toward more complicated models, e.g., with (completely known) interconnected LTI agents \cite{OhMooreAhn-TAC-2014}, and noninterconnected agents subject to local (agent-level) modeling uncertainties \cite{AiYuJiaShenMaYang-RNC-2017}. 

In parallel to the above studies, \cite{RiegerMooreBaldwin-CEIT-2013} articulated the concept of a multilayer control structure according to a graph theoretic consensus viewpoint. It provides an appropriate foundation to study the control aspects of the increasingly important CPSs \cite{Antsaklis-TAC-2014}. Based on a completely known MAS of interconnected single integrators, \cite{Egerstedt-CPS-2015} reported a graph theoretic formulation to capture the architectural aspect of CPSs. Nevertheless, a CPS might be subject to various cyber and physical abnormalities to be addressed using a mixed control and cybersecurity framework (e.g., see \cite{CardenasAminSastry-HTS-2008} and \cite{ChongSandbergTeixeira-ECC-2019}).

From a control perspective, after a few preliminary studies, \cite{RezaeiStefanovic-Automatica-2021} proposed a mixed control and graph theoretic framework to stabilize an interconnected MAS in a distributed manner. That design framework, in particular, enables a designer to capture the architectural aspect of (single or multiagent) CPSs with separate agent (physical) and control (cyber) layers. Nevertheless, while robust with respect to the modeling uncertainties, that approach does not guarantee the stability of interconnected MAS in the presence of cyber attacks.

Reference \cite{Rezaei-CDC-2021} developed another distributed stabilization protocol to simultaneously guarantee a level of resiliency against the denial of service (DoS) attacks over the control layer, and robustness with respect to the modeling uncertainties over the agent layer. That method is based on the concept of average dwell time, which is known to be conservative, e.g., according to section IV in \cite{Rezaei-AIAA-2021}. Moreover, it is not straightforward to expand the theoretical side of that work to a distributed DoS scenario, \textcolor{black}{when an attacker persistently blocks (multiple) individual communication links, instead of blocking the control layer communications for short periods of time in a centralized manner \cite{Rezaei-CDC-2021}}. It is possible to reduce the conservatism using a mode-dependent average dwell time formulation \cite{RezaeiStefanovic-ACC-2021} and expand that to a distributed DoS attack scenario. Nevertheless, such purely control-oriented viewpoints mainly are suitable for the resiliency analysis of a readily compromised systems, or are passive or reactive with respect to the attacks (with no reaction to the attacks, or with a conservative reaction only after the time that an attack is detected).

In the literature of computer science, the above issues have been addressed by cyber resilience or agility techniques, which proactively (without any indication of the attack) change the network communication paths in order to make the routing topology unpredictable and, therefore, rescue the attacked traffic \cite{jafarian2013formal}-\cite{bhardwaj2020multipath}. Via such route randomization or multipath routing, it is shown that the these cyber agility techniques raise the network infrastructures' bar against attacks, such as the distributed DoS attack, which is the focus of this paper. Such a moving target defense concept recently has attracted interest in the control systems community. We refer to \cite{GriffioenWeerakkodySinopoli-TAC-2021} and \cite{KanellopoulosVamvoudakis-TAC-2020} for single dynamical systems. In particular, focusing on a control design problem, \cite{KanellopoulosVamvoudakis-TAC-2020} relies on the redundancy in the physical components of the underlying system and its moving target defense idea is based on an average dwell time condition, which would be conservative (see the previous paragraph). Also, \cite{GiraldoCardenas-Springer-2019} relies on the closed-loop system matrices of the completely known LTI (single dynamical) systems, and its MAS-related analysis is based on the decoupled second-order agents, which will not be applicable to the design problem for a more general MAS in Section~\ref{Section:Main} (with high-order interconnected agents subject to the nonlinear modeling uncertainties).

We propose a new design framework that synergistically combine the cybersecurity and control algorithms in order to effectively handle the multitude of cyber and physical challenges for interconnected MASs. Namely, for a fully heterogeneous interconnected MAS, we consider the modeling uncertainties and nonvanishing perturbations over the agent layer, and distributed DoS attacks over the control layer. From a control viewpoint, we broaden the design aspect to an arbitrarily fast switched distributed stabilization protocol, while capturing the architectural aspect of CPSs (to separately study the cybersecurity concerns). From a cybersecurity viewpoint, we propose a control-aware satisfiability modulo theory (SMT) formulation to develop a set of control layer (communication) subgraphs that satisfy multiple security constraints. We further rely on the arbitrarily fast switching capability of our design in order to secure the control layer communication against distributed DoS attacks, via a highly frequent mutation between the alternative control sublayers. In particular, unlike \cite{Rezaei-CDC-2021} and \cite{FengTesiDePersis-CDC-2017}, we do not restrict the class of (persistent) DoS attacks by any average dwell time means. To validate the feasibility of the proposed proactive cyber defense strategy, we theoretically prove the robust input-to-state stability (ISS) of the two-layer interconnected MAS (with its cybersecurity-aware control sublayer topologies) and, in simulation, discuss its effectiveness in the presence of the distributed DoS attacks.

In the rest of this paper, we overview a few definitions (Section~\ref{Section:Preliminaries}), propose the main results (Section~\ref{Section:Main}), discuss a simulation study (Section~\ref{Section:Simulation}), and summarize the paper (Section~\ref{Section:Summary}).

\section{Preliminaries}\label{Section:Preliminaries}
We use $\textbf{0}$ to denote a matrix of all zeros with compatible dimension, $\text{diag}\{.\}$ a (block) diagonal matrix of the elements in $\{.\}$, $\text{col}\{x_i\}$ an ordered column vector of $x_i\in \mathbb{R}^n$ for a set of $i$, $\|.\|$ the (induced) 2-norm of its input vector (matrix), $\otimes$ Kronecker product, and $A \succ B$ ($\bar{V} \succ 0$) a positive definite matrix $A-B \in \mathbb{R}^{n \times n}$ (scalar function $\bar{V}$). Also, $\succcurlyeq$ indicates positive semi-definiteness. We propose two graph topologies $\mathcal{G}_a$ and $\mathcal{G}_c$ which admit selfloops (i.e., an edge that goes out and returns to the same node without passing others), unlike traditional cases \cite{MesbahiEgerstedt-Princeton-2010}, and can be disconnected.

An \textit{agent layer} digraph $\mathcal{G}_a$ with $N$ nodes represents the physical interaction of agents' dynamics. It is characterized by a modified adjacency matrix $\mathcal{A}_a = [a_{ij}^{a}] \in \mathbb{R}^{N \times N}$ where $N \geq 1$ is the total number of agents, $a_{ij}^a \neq 0$ if the $i^{th}$ agent is affected by the $j^{th}$ agent's dynamics for $i,j \in \{1,2,...,N\}$, and $a_{ij}^a = 0$ otherwise. Compared to the standard definition, selfloops $j = i$ are admitted, and both positive and negative signs are acceptable for $a_{ij}^a$. Information about all agent-layer neighbors of an agent $i$, including the $i^{th}$ agent if there is a selfloop, is given by a set $\mathcal{N}_i^a$. Only a scalar $\|\mathcal{A}_a\|$ is shared with the control layer designer, such that the interconnection topology (structure and edge weights) remains confidential. It enables a designer to eliminate the predictability of the control layer topology, to be discussed in the first paragraph of Subsection~\ref{Subsection:DesignAndAnalysis}.

A \textit{control layer} is built by $M \geq 1$ control sublayer(s), to be determined by a piecewise constant switching signal $\sigma\in \{1,2,...,M\}$. Each control sublayer's undirected graph $\mathcal{G}_{c\sigma}$ is represented by a modified Laplacian matrix $\mathcal{H}_{c\sigma} = [h_{ij}^{c \sigma}] \in \mathbb{R}^{N \times N}$ where $i,j \in \{1,2,...,N\}$, $h_{ij}^{c \sigma} = -a_{ij}^{c \sigma}$, and $h_{ii}^{c \sigma} = \sum_{j \in \mathcal{N}_i^{c \sigma}} a_{ij}^{c \sigma} + s_i^{c \sigma}$, $\mathcal{N}_i^{c \sigma}$ denotes the neighbor set of the control node $i$ (without selfloops), $a_{ij}^{c \sigma} \geq 0$ edge weights, and $s_i^{c \sigma} > 0$ selfloop weights of control node $i$ ($s_i^{c \sigma}= 0$ if there is not a selfloop). Each (fixed) control sublayer topology $\mathcal{G}_c^\sigma$ can be disconnected; however, there is at least one selfloop in each connected component of $\mathcal{G}_c^\sigma$. Consequently, all eigenvalues of $\mathcal{H}_{c \sigma}$ are real-valued positive scalars to be sorted as $0 < \mu_{\sigma 1} \leq \mu_{\sigma 2} \leq ... \leq \mu_{\sigma N}$ \cite{RezaeiStefanovic-Automatica-2021}. We use \textit{both structure and weights} (i.e., \textit{topology}) of each subgraph $\mathcal{G}_{c \sigma}$ as the design degrees of freedom in Subsections~\ref{Subsection:DesignAndAnalysis} and~\ref{Subsection:MDT}. In particular, $\mathcal{G}_{c \sigma}^{01}$ abstracts the structure of $\mathcal{G}_{c \sigma}$, to be determined using the cybersecurity Algorithms \ref{alg:main} and~\ref{alg:smt}.

We let $\{t_k\}_{k \in \mathbb{Z}_{\geq 0}}$ be the switching time sequence for $\sigma(t)$, and assume that there exists a dwell time $\tau_d>0$ such that multiple switchings will not happen at the same time: $\inf_{k \in \mathbb{Z}_{\geq 0}} (t_{k+1} - t_k) \geq \tau_d$, the adjacent switching intervals $[t_k, t_{k+1})$ do not have overlaps, and the control sublayer $\mathcal{G}_{c \sigma}$ remains fix during each switching time interval. Unlike \cite{Rezaei-AIAA-2021} and \cite{RezaeiStefanovic-ACC-2021}, we neither restrict the length of $\tau_d$ by any means nor use it in the derivations of this paper.

Over a control layer, selfloops determine the control layer configuration (see subsection 3.1 in \cite{RezaeiStefanovic-IFAC-2020}). Over an agent layer, selfloops model local or agent-level modeling uncertainty when an agent's modeling uncertainty depends on its own state variables (see the description of agents~\eqref{eq:AgentModel}).

\section{Main Results}\label{Section:Main}
In this section, we lay the foundation of this paper (Subsection~\ref{Subsection:ProblemStatement}), develop a new framework to design and validate a control layer (Subsections~\ref{Subsection:DesignAndAnalysis} and~\ref{Subsection:MDT}), and provide a theoretical analysis for the proposed two-layer (closed-loop) interconnected MAS (Subsection~\ref{Subsection:TheoreticalAnalysis}).

\subsection{Problem Foundation}\label{Subsection:ProblemStatement}
We consider an MAS of $N$ interconnected agents:
\begin{equation}\label{eq:AgentModel}
	\begin{array}{rl}
		\dot{x}_i(t) & = A_i x_i(t) + B_{u_i} u_i(t) + B_{f_i} f_i(y_i(t),t) + B_{d_i} d_i(t)\\
		        y_i(t) & = C_{y_i} \sum_{j \in \mathcal{N}_i^a} a_{ij}^a x_j(t)
	\end{array}
\end{equation}
\normalsize
where $i \in \{1,2,...,N\}$ denotes the agent number; $x_i \in \mathbb{R}^{n_x}$ state variable of agent $i$; $u_i \in \mathbb{R}^{n_u}$ control input; $d_i \in \mathbb{R}^{n_d}$ nonmeasurable, nonvanishing, external perturbation (disturbance or process noise); and $y_i \in \mathbb{R}^{n_y}$ interconnection variable. The known $A_i \in \mathbb{R}^{n_x \times n_x}$ is the $i^{th}$ agent's system matrix, $B_{u_i} \in \mathbb{R}^{n_x \times n_u}$ control allocation matrix, $B_{d_i} \in \mathbb{R}^{n_x \times n_d}$ perturbation allocation matrix, and $B_{f_i} \in \mathbb{R}^{n_x \times n_g}$ uncertainty allocation matrix. Each pair $(A_i, B_{u_i})$ represents a stabilizable system, and $B_{d_i}$ and $B_{f_i}$ can be in the range space of $B_{u_i}$ (matched scenario) or not (unmatched scenario). This enables a designer to model an MAS subject to the mixed matched and unmatched modeling uncertainties. 

The unknown interconnection matrices $C_{y_i} \in \mathbb{R}^{n_y \times n_x}$ satisfy the norm conditions $\|C_{y_i}\|^2 \leq \gamma_{cyi}$. The nonlinear functions $f_i: \mathbb{R}^{n_y} \times \mathbb{R}_{\geq 0} \to \mathbb{R}^{n_g}$ satisfy Lipschitz condition to ensure the existence and uniqueness of the solutions to the nonlinear differential equations~\eqref{eq:AgentModel} \cite{Khalil-Prentice-2003}. To avoid a conservative Lipschitz-based stabilization approach, we assume that each nonlinear function satisfies a norm condition $f_i^T(y_i,t) f_i(y_i,t) \leq \gamma_{f_i} y_i^T y_i$. We also assume that only two constants, $\gamma_{cy} = \max_i (\gamma_{cyi}) >0$ and $\gamma_f =  \max_i \{\gamma_{f_i}\}>0$, are known to the control layer designer. (A Lipschitz constant satisfies this norm condition; however, it would end in a larger (more conservative) constant compared to $\gamma_f$.)\vspace{0.04in}

Model~\eqref{eq:AgentModel} enables us to consider an interconnected MAS subject to both agent-level modeling uncertainties (via selfloops with unknown $a_{ii}^a$ over the agent layer: $i \in \mathcal{N}_i^a$) and MAS-level modeling uncertainties (via non-selfloop edges with unknown $a_{ij}^a$: $j \neq i$ and $ j \in \mathcal{N}_i^a$), because the agents' neighbor sets $\mathcal{N}_i^a$ (over agent layer) are unknown.\vspace{0.04in}

We propose the following communication-based, switched distributed stabilization protocol:
\begin{equation}\label{eq:StabilizationProtocol}
	u_i(t) = \big(\sum_{j \in \mathcal{N}_i^{c\sigma(t)}} a_{ij}^{c\sigma(t)} (v_i(t) - v_j(t)) + s_i^{c\sigma(t)} v_i(t) \big)
\end{equation}
\normalsize
in which each virtual stabilization signal $v_i$ is locally computed by the associated agent $i$:
\begin{equation}\label{eq:virtualstabilziation}
	v_i(t) = K_i x_i(t)
\end{equation}
\normalsize
where $K_i \in \mathbb{R}^{n_u \times n_x}$ denotes the $i^{th}$ agent's stabilization gain, to be designed.\vspace{0.04in}

We aggregate the interconnected agents~\eqref{eq:AgentModel} for all $i \in \{1,2,...,N\}$ to find a model of the agent layer:
\begin{equation}\label{eq:AgentLayer}
\begin{array}{rl}
	\dot{x}(t) & = \bar{A} x(t) + \bar{B}_u u(t) + \bar{B}_f f(y,t) + \bar{B}_d d(t)\\
			y(t) & = \bar{C}_y (\mathcal{A}_a \otimes I_{n_x}) x(t)~~~
\end{array}
\end{equation}
\normalsize
where $\bar{A} := \text{diag}\{A_i\}$, $\bar{B}_u := \text{diag}\{B_{u_i}\}$, $\bar{B}_f := \text{diag}\{B_{f_i}\}$, $\bar{B}_d := \text{diag}\{B_{d_i}\}$, $\bar{C}_y := \text{diag}\{C_{y_i}\}$, $x := \text{col}\{x_i\}$, $u := \text{col}\{u_i\}$, $d := \text{col}\{d_i\}$, $y := \text{col}\{y_i\}$, and $f := \text{col}\{f_i\}$. We also aggregate the distributed stabilization protocols~\eqref{eq:StabilizationProtocol} for all $i$, and reach to the following model of the switched control layer (with multiple sublayers):
\begin{equation}\label{eq:ControlLayer}
	u(t) = (\mathcal{H}_{c \sigma(t)} \otimes I_{n_u}) v(t)
\end{equation}
\normalsize
where $v := \text{col}\{v_i\} := \bar{K} x$, and $\bar{K} := \text{diag}\{K_i\}$. Now we are ready to articulate the main objective of this paper. (See \cite{Sontag-NOC-2008} for the definitions of comparison functions and ISS.)

\noindent\rule{\columnwidth}{1pt}
\textbf{Objective}: In the presence of modeling uncertainties and nonvanishing perturbations (over agent layer) and distributed DoS attacks (over control layer), design a distributed protocol to guarantee both robust ISS and proactive security for a two-layer (closed-loop) interconnected MAS \eqref{eq:AgentLayer} and \eqref{eq:ControlLayer}:
\begin{IEEEeqnarray}{rcl}
		\|x(t)\| \leq \beta(\|x(0)\|,t) + \kappa(\|d(t)\|_\infty)
\end{IEEEeqnarray}
\normalsize
for $\mathscr{KL}$-function $\beta$, $\mathscr{K}_\infty$-function $\kappa$, and $d \in L_\infty(\mathbb{R}_{\geq 0})$.\vspace{0.03in}

\noindent\rule{\columnwidth}{1pt}

Modified based on \cite{Sontag-NOC-2008} and~\cite{Liberzon-Springer-2003}, we also define a common ISS Lyapunov function to prove robust distributed ISS in Subsection~\ref{Subsection:TheoreticalAnalysis}.\vspace{0.05in}

\begin{definition}\label{Definition:ISS-Lyapunov-Common}
	 A function $\bar{V}: \mathbb{R}^{n} \to \mathbb{R}_{\geq 0}$ is a smooth, common ISS Lyapunov function for the two-layer interconnected MAS \eqref{eq:AgentLayer} and \eqref{eq:ControlLayer} if the following inequalities are satisfied:
\begin{IEEEeqnarray*}{rLL}
	\kappa_1(\|x\|) \leq \bar{V}(x) & \leq \kappa_2(\|x\|) \\
	\dot{\bar{V}}(x) & \leq - \kappa_3(\|x\|) + \gamma(\|d\|)
\end{IEEEeqnarray*}
for all $\sigma \in \{1,2,...,M\}$, and for $\mathscr{K}_\infty$ functions $\kappa_1$, $\kappa_2$, $\kappa_3$, and $\gamma$. \hfill $\blacktriangleleft$\vspace{0.03in}
\end{definition}

\subsection{Control Layer Design Framework: Stabilization}\label{Subsection:DesignAndAnalysis}
In order to go beyond the ideas of \cite{Rezaei-CDC-2021} and establish proactive cyber defense, we propose a design procedure to simultaneously:
\begin{enumerate}
	\item make the control layer (communication) topology independent of the agent layer (interconnection) topology, i.e., to capture the architectural aspect of CPSs, and
	\item reconfigure the control layer topology in an arbitrarily fast manner.
\end{enumerate}

Note that at least a subset of communication links would be predictable when a known agent layer topology is included in the control layer topology \cite{FengTesiDePersis-CDC-2017}, or if we follow the (average dwell time-based) resiliency bounds in~\cite{Rezaei-CDC-2021} and create a slow-varying switching signal. Instead, we propose the following design procedure to develop a control layer with an arbitrarily fast switching of multiple sublayers (Modified from \cite{RezaeiStefanovic-ACC-2021} and \cite{RezaeiStefanovic-MED-2021} to develop a ``common" ISS Lyapunov function in Subsection~\ref{Subsection:TheoreticalAnalysis}).\vspace{0.025in}

\noindent\rule{\columnwidth}{1pt}
\begin{design}\label{DesignProcedure}(A mixed optimal control and graph theoretic formulation)~\\
	\begin{enumerate}
		\item For each control sublayer $\sigma \in \{1,2,...,M\}$, develop a control layer graph topology $\mathcal{G}_{c\sigma}$ based on a {\sl cooperative configuration}: Set a few $s_i^{c\sigma} > 0$ to define selfloops, and assign $\mathcal{N}_i^{c\sigma}$ and $a_{ij}^{c\sigma(t)}>0$ to a graph in which each connected component has a control node with a selfloop. Let $\mu_{min} := \min_\sigma \{\mu_{\sigma 1}\} > 0$ where $\mu_{\sigma 1}$ is the smallest (positive) eigenvalue of $\mathcal{H}_{c \sigma}$ associated to each $\mathcal{G}_{c \sigma}$.
		\item For each agent $i \in \{1,2,...,N\}$, design a candidate robust ISS gain $K_i$ as follows:
	\begin{enumerate}
		\item Let $Q_i \in \mathbb{R}^{n_x \times n_x}$ and $R_i \in \mathbb{R}^{n_u \times n_u}$ be two positive definite design matrices, and $Q_{f i} = Q_i + R_f$ a modified state weighting matrix where $R_f = (a_f \gamma_f \gamma_{cy} \|\mathcal{A}_a\|^2 + a_d) I_{n_x}$ for two positive design scalars $a_f$ and $a_d$.\vspace{0.05in}
		\item Find the optimal solution $v_i' = K_i x_i' \in \mathbb{R}^{n_u}$ of the following modified linear quadratic regulator (LQR) problem:
	\begin{equation*}
		\begin{matrix}
			\min_{v_i' \in \mathcal{C}_{i}} & \int_0^\infty (x_i'^T Q_{f i} x_i' + v_i'^T R_i v_i')d\tau\\
			\textrm{subject~to} & \dot{x}_i' = A_i x_i' + \mu_{min} B_{u_i} v_i'
		\end{matrix}
	\end{equation*}
	\normalsize 
	in which $\mathcal{C}_{i}$ is the set of all static linear state feedback stabilizing signals $v_i'$ for the networked nominal dynamics $\dot{x}_i' = A_{i} x_i' + \mu_{min} B_{u_i} v_i'$.
	 \end{enumerate}
		\item $\mathcal{G}_{c \sigma}$ and $K_i$ build a set of valid control sublayers if the following condition is satisfied:
		\begin{IEEEeqnarray}{rll}\label{eq:DesignCondition}
			\bar{Q}_{v \sigma} := \bar{Q} + \bar{K}^T \big( \bar{R} + 2\bar{R} \bar{E}_{c \sigma} \big) \bar{K} - \frac{1}{a_f} \bar{P} \bar{B}_f \bar{B}_f^T \bar{P} \succ \textbf{0} ~~~~~ 
		\end{IEEEeqnarray}
		\normalsize
		and $\bar{Q} := \text{diag}\{Q_i\}$, $\bar{R} :=  \text{diag}\{R_i\}$, $\bar{P} := \text{diag}\{P_i\}$, and $\bar{E}_{c \sigma} := \big((\frac{\mathcal{H}_{c \sigma}}{\mu_{min}} - I_N) \otimes I_{n_u}\big) \succcurlyeq \textbf{0}$. The matrices $P_i \in \mathbb{R}^{n_x \times n_x}$ are the unique positive definite solutions of the following algebraic Riccati equations (AREs):
	\begin{equation}\label{eq:ARE}
		A_i^T P_i + P_i A_i + Q_{f i} - \mu_{min}^2 P_i B_{u_i} R_i^{-1} B_{u_i}^T P_i = \textbf{0}.
	\end{equation}
	\normalsize
   \end{enumerate}
\end{design}
\noindent\rule{\columnwidth}{1pt}
\vspace{0.025in}

The term ``modified" highlights the required modifications to obtain a state weighting matrix $Q_{fi}$ and the presence of $\mu_{min}$ in the networked nominal dynamics of the modified LQR formulation. The existence and uniqueness of the solutions $P_i$ are guaranteed by the stabilizability and observability of the triple $((Q_{f i})^{1/2}, A_{i}, \mu_{min} B_{u_i})$ where $((Q_{f i})^{1/2})^T (Q_{f i})^{1/2} = Q_{f i}$, pointing out that each $(A_{i}, \mu_{min} B_{u_i})$ is stabilizable due to that of $(A_i,B_{u_i})$ and positiveness of $\mu_{min}$. We recommend to develop a set of control sublayers $\mathcal{G}_{c \sigma}$ (or $\mathcal{H}_{c \sigma}$) with sufficiently positive $\mu_{min}$ in order to avoid controllability issues in the modified LQR problems (or, equivalently, singularity in AREs~\eqref{eq:ARE}).\vspace{0.05in}

\begin{remark}
    Arbitrarily fast switching has already been used in the literature of MASs. However, to the best of our knowledge, the existing studies should be limited to the consensus (vs. stabilization) problem for completely known, homogeneous, or purely LTI MASs with noninterconnected dynamics (e.g., see \cite{ValcherZorzan-Automatica-2017}). Such developments are not necessarily applicable to the considered problem in this paper.\vspace{0.05in}\hfill $\blacktriangleleft$
\end{remark}
\subsection{Control Layer Design Framework: Cybersecurity}\label{Subsection:MDT}
In the previous subsections, we proposed a switched distributed stabilization protocol~\eqref{eq:StabilizationProtocol} with all stabilization gains and control sublayer topologies as the design degrees of freedom, and developed a step-by-step procedure to design and validate a set of control sublayers for the robust ISS of a two-layer interconnected MAS~\eqref{eq:AgentLayer} and~\eqref{eq:ControlLayer} with an arbitrarily fast switching. Together with a control-oriented recommendation at the end of the previous subsection, the Step~1 of that procedure gives insights to determine a control layer configuration from a large-scale system viewpoint~\cite{Lunze-Prentice-1992}. To better leverage the power of the proposed distributed stabilization protocol~\eqref{eq:StabilizationProtocol}, along the Step~1 of Design Procedure~\ref{DesignProcedure}, we propose a new formulation to determine a set of candidate control sublayer structures that simultaneously satisfy multiple cybersecurity and distributed control constrains:
\begin{enumerate}
    \item {\sl Full connectivity}: To ensure that the conditions of the control subgraphs in Section~\ref{Section:Preliminaries} are satisfied, \textcolor{black}{while taking into account the agents' capability or willingness to provide their absolute measurements to the control layer (e.g., see subsection~3.1 and section 5 in~\cite{RezaeiStefanovic-IFAC-2020})}.
    \item {\sl \textcolor{black}{Centrality} distribution}: To have a sufficient number of active selfloops to reduce attack vulnerability by avoiding a single point of failure over the control layer.
    \item {\sl Non-overlapping paths}: To ensure that an attacker will not be able to compromise multiple control sublayers by a single attack on a common inter-controller communication link (i.e., \textcolor{black}{to increase the cost of attack}).
    \item {\emph{Low risk}}: By excluding the high risk communication links that have been compromised in the recent past.\vspace{0.025in}
\end{enumerate}

Note that these conditions would naturally disqualify centralized (with all-to-all) and decentralized (with only selfloops) configurations from the Step~1 of Design Procedure~\ref{DesignProcedure}. Further, the above planning is a generalization of the 0-1 knapsack problem \cite{sahni1975approximate}, which is NP-hard. Therefore, we \textcolor{black}{reformulate} it as a satisfiability problem using a generalized Boolean/arithmetic logic of SMT \cite{BM09}. 
(While satisfiability problems are NP-complete in general, the recent advances in SMT solvers have made them scalable to the problems with millions of variables \cite{Moura2009a}.) 

In conjunction with the Step 1 of Design Procedure~\ref{DesignProcedure}, we propose Algorithm \ref{alg:main} to generate a set $\mathscr{G}_{c \sigma}^{01}$ of $M$ control sublayer structures $\mathcal{G}_{c \sigma}^{01} = (\mathcal{V}, \mathcal{E}_{c\sigma}^{01})$, where $\mathcal{V} = \{1, 2, \ldots, N\}$ denotes the node set and $\mathcal{E}_{c\sigma}^{01} = \{\alpha_{ij}^{c\sigma} ~~ \forall ~~ (i, j) \in \mathcal{V} \times \mathcal{V}\}$ edge set over each $\mathcal{G}_{c \sigma}^{01}$. The edge weights $\alpha_{ij}^{c\sigma}$ of this outcome graph are either $1$ when an edge is determined by the proposed algorithms, or $0$ otherwise. Further, to introduce fewer parameters in this subsection, we use $\alpha_{ii}^{c\sigma}$ to denote the existence of a selfloop around any node $i$. These two points are unlike the definitions of (real-valued) $a_{ij}^{c \sigma} \in \mathbb{R}_{\geq 0}$ and $s_i^{c\sigma} \in \mathbb{R}_{\geq 0}$ for the final $\mathcal{G}_{c \sigma}$ (see Section~\ref{Section:Preliminaries}). An integer-valued $T$ denotes the number of the nodes that are required to have selfloops (i.e., for the centrality distribution). Each integer-valued scalar $\theta_i \in \{0,1\}$ determines whether we can add a selfloop to a node $i$ and use the absolute measurement of that agent, or not. Each integer-valued scalar $r_{ij} \in \{0,1\}$ indicates whether a link $(i,j)$ \textcolor{black}{(therefore, $(j,i)$)} is a high risk link or not. Further, each integer-valued scalar $\eta_{ij} \in \{0,1\}$ memorizes whether a link $(i,j)$ \textcolor{black}{(therefore, $(j,i)$)} has been inactive in all prior control sublayer graphs or not, i.e., an $\eta_{ij} = 1$ means that link has not been used previously.\vspace{0.05in}

\begin{algorithm}[b!]
\caption{$generateAllGraphs$ \label{alg:main}}
\textbf{Inputs}: $M$, $N$, $T$, $\{\theta_i \}$, $\{r_{ij} \}$\\
\textbf{Output:} $\mathscr{G}_{c \sigma}^{01}$\\
\CommentSty{\%initialization}\\
    $\mathcal{V} = \{ 1, \ldots, N\}$\\
    \For{\text{every} $(i,j) \in \mathcal{V} \times \mathcal{V}$}
    {
       $\eta_{ij} = 0$
    }
    
\CommentSty{\%Generate all (sub) graphs}

    \For{$\sigma = 1$ to $M$}
    {
        
        $\mathcal{G}_{c \sigma}^{01}$ = generateOneGraph($\mathcal{V}, T, \{r_{ij}\}, \{\theta_i\}, \{\eta_{ij} \}, \sigma$)

        $\mathscr{G}_{c \sigma}^{01}  = \mathscr{G}_{c \sigma}^{01} \cup \mathcal{G}_{c \sigma}^{01}$
        
        \For{every $(i,j) \in \mathcal{V} \times \mathcal{V}, i \neq j$}
        {
            $\eta_{ij} = \eta_{ij} \lor \neg\alpha_{ij}^{c\sigma}$ 
        }
    }
    \normalsize
\end{algorithm}

Based on a logic (notation) similar to the standard SMT references, e.g., \cite{BM09}, Algorithm \ref{alg:smt} describes the proposed SMT formulation to generate a candidate control sublayer structure $\mathcal{G}_{c\sigma}^{01}$ that satisfies the aforementioned user-defined conditions 1 to 4. {\color{black} In particular, we first consider a directed (sub) graph design problem, in order to determine the communication paths over each control sublayer. {\sl Constraint~\eqref{eq:alpha_bound}} ensures the soundness of the resulting control layer subgraphs, by limiting all integer-valued $\alpha_{ij}^{c\sigma}$ to either $0$ or $1$. {\sl Constraint~\eqref{eq:can_selfloop}} ensures only the nodes with selfloop capability \textcolor{black}{will be asked to share their absolute measurements over the control sublayers} (see subsection 3.1 in \cite{RezaeiStefanovic-IFAC-2020}). {\sl Constraint~\eqref{eq:num_of_selfloops}} ensures that $T$ nodes in the outcome sublayer graph will have active selfloops (to avoid a single point of failure). {\sl Constraint~\eqref{eq:complete_change}} ensures that all the \textcolor{black}{high-risk, recently compromised communication links (known based on the history of the underlying two-layer interconnected MAS)} will be excluded from the control layer. {\sl Constraint~\eqref{eq:risk_free}} ensures that any link which has been used in the prior control sublayers \textcolor{black}{(developed by the proposed algorithm)} will not be used in the new sublayer graph. {\sl Constraint~\eqref{eq:connectivity_main}} ensures that any node without an active selfloop will be connected to a node with an active selfloop (either directly or via an intermediary node). {\sl Constraint~\eqref{eq:exactly_one_connected}} ensures that every node has either an active selfloop or has a path to \textcolor{black}{(receive information from)} {exactly} one node with an active selfloop. A solver $solveSMTModel(.)$ (e.g., see \cite{BM09}) determines the satisfiable assignments to the unknown variables for an SMT model $\Lambda$, in order to obtain the active communication links. \textcolor{black}{Finally, an edge set $\mathcal{E}_{c\sigma}^{01}$ is built, in order to offer the structure of a symmetric control sublayer $\mathcal{G}_{c \sigma}^{01}$ with undirected communication links.} \vspace{0.025in}}


\begin{algorithm} [t!]
\small
\caption{$generateOneGraph$ \label{alg:smt}}
\textbf{Inputs:} $\mathcal{V}, T, \{r_{ij}\}, \{\theta_i\}, \{\eta_{ij} \}, \sigma$\\
\textbf{Output: $\mathcal{G}_{c \sigma}^{01}$}\\
\CommentSty{\%Define SMT model $\Lambda$, with following constraints}\\
\begin{align}
& 0 \leq \alpha_{ij}^{c\sigma} \leq \ 1, & \forall (i,j) \in \mathcal{V} \times \mathcal{V} \label{eq:alpha_bound} 
\end{align}
\begin{align}
& \alpha_{ii}^{c\sigma} \leq \theta_i, & \forall i \in \mathcal{V} \label{eq:can_selfloop} \\
& \sum_{i \in \mathcal{V}} \alpha_{ii}^{c\sigma} = T  \label{eq:num_of_selfloops}
\end{align}
\begin{align}
& \alpha_{ij}^{c\sigma} \leq r_{ij} & \forall (i,j) \in \mathcal{V} \times \mathcal{V} \label{eq:complete_change} \\
& \alpha_{ij}^{c\sigma} \leq \eta_{ij}, &  \forall (i,j) \in \mathcal{V} \times \mathcal{V} \label{eq:risk_free}
\end{align}
\begin{align}
& (\alpha_{ij}^{c\sigma} = 1) \land (\alpha_{jk}^{c\sigma} = 1) \rightarrow (\alpha_{kk}^{c\sigma} = 1) \land (\alpha_{jj}^{c\sigma} = 0) \nonumber \\
& \forall i, \forall j \neq i, \forall k \neq j \label{eq:connectivity_main}
\end{align}
\begin{align}
& \sum_{j \in \mathcal{V}} \alpha_{ij}^{c\sigma} = 1, & \forall i \in \mathcal{V} \label{eq:exactly_one_connected}
\end{align}

\CommentSty{\%Solve the SMT model}

$\{\alpha_{ij}^{c\sigma}\}$ = solveSMTModel$(\Lambda)$

\CommentSty{\%\textcolor{black}{Build a symmetric (sub) graph structure}}

$\alpha_{ji}^{c\sigma} = 1, ~~~~~ \forall \alpha_{ij}^{c\sigma} = 1$

$\mathcal{E}_{c\sigma}^{01} = \{\alpha_{ij}^{c\sigma} ~~ \forall (i, j) \in \mathcal{V} \times \mathcal{V}\}$

$\mathcal{G}_{c \sigma}^{01} = (\mathcal{V}, \mathcal{E}_{c\sigma}^{01})$
\normalsize
\end{algorithm}

The exclusion of high risk links enhances the resilience of the control layer to the potential distributed DoS attacks. However, we ensure the cyber agility via a consistent mutation of the control sublayers (see Design Procedure~\ref{DesignProcedure}, and Algorithms~\ref{alg:main} and~\ref{alg:smt}) based on an arbitrarily fast switching strategy. Such an arbitrarily fast and unordered switching would turn a control layer's communication links (or topology) into a set of unpredictable moving targets. Next to the fact that the candidate sublayers have no common inter-controller communication link, \textit{1)} it would be hard(er) for an attacker to study the underlying two-layer interconnected MAS in order to plan and execute a distributed DoS attack, and \textit{2)} a distributed DoS attack that targets a specific set of inter-controller communications in any one of these candidate sublayers, will not be effective against others.

{\color{black} We need to mention that the proposed SMT formulation is based on the notion of directed subgraphs, which would end in the structurally nonsymmetric control sublayers (see \cite{Rezaei-AIAA-2021} and \cite{RezaeiStefanovic-IFAC-2020}). We manually set $\alpha_{ji}^{c\sigma} = 1$ if the SMT solver assigns $\alpha_{ij}^{c\sigma} = 1$ to a pair $(i, j) \in \mathcal{V} \times \mathcal{V}$, in order to build a control sublayer graph with undirected communication links. This viewpoint is along the practices in computer science, where the computer networks' communication links are full-duplex. We also need to mention that the proposed SMT formulation determines the structure $\mathcal{G}_{c\sigma}^{01}$ of a control sublayer topology $\mathcal{G}_{c \sigma}$. We manually select the weights of edges and selfloops, and design a set of distributed stabilization gains, such that the validation condition~\eqref{eq:DesignCondition} is satisfied. Expanding the high-dimension approach in~\cite{RezaeiStefanovic-RNC-2019} (limited to the first and second order agents) to the agent model~\eqref{eq:AgentModel}, an interested reader might be able to combine that high-dimension modified LQR formulation with the low-dimension one in Step 2 of Design Procedure~\ref{DesignProcedure} in order to automatically determine a set of robust edge weights and selfloops based on the subgraph structures of Algorithms~\ref{alg:main} and~\ref{alg:smt}.}


\subsection{Theoretical Analysis}\label{Subsection:TheoreticalAnalysis}
In this subsection, we derive a few key properties of the proposed design framework, and analyze ISS for the resulting two-layer interconnected MAS~\eqref{eq:AgentLayer} and~\eqref{eq:ControlLayer} with a security-oriented, arbitrarily fast switching of the control sublayers.

Analytically, we know that the stabilization gains $K_i$ are characterized as follows \cite{Lin-Wiley-2007}:
\begin{IEEEeqnarray}{rll}\label{eq:ControlGain}
	\begin{array}{rl}
		K_i & = -\mu_{min} R_i^{-1} B_{u_i}^T P_i.
	\end{array}
\end{IEEEeqnarray}
\normalsize

We aggregate each \eqref{eq:ARE} and \eqref{eq:ControlGain}, and further find $\bar{A}^T \bar{P} + \bar{P} \bar{A} + \bar{Q}_f - \mu_{min}^2 \bar{P} \bar{B}_u \bar{R}^{-1} \bar{B}_u^T \bar{P}  = \textbf{0}$ and $\bar{K} + \mu_{min} \bar{R}^{-1} \bar{B}_u^T \bar{P} = \textbf{0}$ where $\bar{Q}_f := \text{diag}\{Q_{f i}\}$. We postmultiply both of the above equalities by $x$, premultiply the second one by $x^T$, and after a few manipulations find the following design properties for each fixed control sublayer (known as the optimality conditions in the literature of optimal control for single dynamical systems \cite{Lin-Wiley-2007}).\vspace{0.06in} 

\begin{fact}\label{FactUM}
	The following equalities hold for each control sublayer of Design Procedure~\ref{DesignProcedure} and Algorithms~\ref{alg:main} and~\ref{alg:smt}, to be used in a two-layer interconnected MAS \eqref{eq:AgentLayer} and \eqref{eq:ControlLayer}:
	\begin{IEEEeqnarray*}{rll}
	    2 v^T \bar{R} + \mu_{min} \bar{V}_{x}^T \bar{B}_u = & \textbf{0} \\
    		x^T \bar{Q}_f x + v^T \bar{R} v + \bar{V}_{x}^T (\bar{A} x + \mu_{min} \bar{B}_u v) = & 0.
    \end{IEEEeqnarray*}
    \normalsize
    where $V_{x}^T := \frac{\partial \bar{V}}{\partial x}$ and $\bar{V}(x) := x^T \bar{P}x$.\vspace{0.06in}  \hfill$\blacktriangleleft$
\end{fact}

Now we propose the main result of this subsection.\vspace{0.05in}

\begin{theorem}\label{Theorem:Main}
	If $\mathcal{G}_{c \sigma}$ and $K_i$ are developed according to Subsections~\ref{Subsection:DesignAndAnalysis} and~\ref{Subsection:MDT}, robust ISS is guaranteed for the interconnected MAS~\eqref{eq:AgentLayer} and~\eqref{eq:ControlLayer} despite the modeling uncertainties and nonvanishing perturbations over the agent layer, and arbitrarily fast switching of the control layer (communication) topology.\vspace{0.04in}
\end{theorem}

\textit{Proof}: To facilitate the derivations of this proof, we first substitute the control layer~\eqref{eq:ControlLayer} in the agent layer dynamics~\eqref{eq:AgentLayer}, add and subtract $\mu_{min} \bar{B}_u v$, and rewrite the two-layer interconnected MAS of this paper as follows:
\begin{equation}\label{eq:MASAggregated2}
		\underbrace{\dot{x} = \bar{A} x + \mu_{min} \bar{B}_u v}_{\text{Networked nominal dynamics}} + \underbrace{\bar{B}_f f(y) + \mu_{min} \bar{B}_u \bar{E}_{c\sigma} v + \bar{B}_d d}_{\text{Uncertainties and perturbations over~} \mathcal{G}_a~\text{and}~\mathcal{G}_c}
\end{equation}
\normalsize
in which $\bar{B}_d d$ represents the nonvanishing perturbations and $\bar{B}_f f(z)$ the actual modeling uncertainties over the agent layer, and $\mu_{min} \bar{B}_u \bar{E}_{c \sigma} v$ a fictitious uncertainty over the control layer, that is introduced to formulate a low-dimension modified LQR problem in Design Procedure~\ref{DesignProcedure}, despite the fact that we are dealing with a high-dimension (or MAS-level) robust ISS problem.  \vspace{0.04in}

\noindent We define a candidate common ISS Lyapunov function:
\begin{IEEEeqnarray*}{rll}
	\bar{V}(x) = x^T \bar{P} x \succ 0
\end{IEEEeqnarray*}
\normalsize
which is the same as that of Design Properties~\ref{FactUM} (a consequence of the mixed optimal and graph theoretic formulation in Design Procedure~\ref{DesignProcedure}). As a key point, unlike the multiple Lyapunov functions in \cite{Rezaei-AIAA-2021} and \cite{RezaeiStefanovic-ACC-2021}, we point out that this $\bar{P}$ does not vary depending on the active control sublayer (or switching mode $\sigma$).\vspace{0.04in}

\noindent Along the uncertain trajectories of the two-layer interconnected MAS~\eqref{eq:MASAggregated2} and based on Design Properties~\ref{FactUM}, we find:
	\begin{equation*}
		\begin{array}{rl}
			\dot{\bar{V}} = &  \bar{V}_x^T \dot{x}\\
							   = &  \bar{V}_x^T (\bar{A} x + \mu_{min} \bar{B}_u v) + \mu_{min} \bar{V}_x^T \bar{B}_u \bar{E}_{c\sigma} v + \bar{V}_x^T  \bar{B}_f f(y) + \bar{V}_x^T \bar{B}_d d\\
							   = & -x^T \bar{Q}_f x - v^T \bar{R} v - 2 v^T \bar{R} \bar{E}_{c\sigma} v + 2x^T \bar{P} \bar{B}_f f(y) + 2x^T \bar{P} \bar{B}_d d.
		\end{array}
	\end{equation*}
	\normalsize

\noindent We use Young's inequality as follows:
	\begin{equation*}
		\begin{array}{rl}
		  \dot{\bar{V}}  \leq & -x^T \bar{Q}_f x - v^T \bar{R} v - 2 v^T \bar{R} \bar{E}_{c\sigma} v + a_f f^T f + a_d x^T x + \frac{1}{a_f} x^T \bar{P} \bar{B}_f \bar{B}_f^T \bar{P} x + \frac{1}{a_d} d^T \bar{B}_d^T \bar{P}^2 \bar{B}_d d\\
						  \leq & -x^T\big( \bar{Q} + \bar{K}^T \bar{R} \bar{K} + 2 \bar{K}^T \bar{R} \bar{E}_{c\sigma} \bar{K} - \frac{1}{a_f}\bar{P} \bar{B}_f \bar{B}_f^T \bar{P} \big) x + \gamma_d \|d\|^2  \leq -x^T \bar{Q}_{v \sigma} x + \gamma_d \|d\|^2\\
						  \leq & -x^T \bar{Q}_v x + \gamma_d \|d\|^2
		\end{array}
	\end{equation*}
	\normalsize
	where $\gamma_d := \| \bar{P} \bar{B}_d \|^2$, and $\textbf{0} \prec \bar{Q}_v \preccurlyeq \bar{Q}_{v \sigma}$ for all $\sigma \in \{1,2,...,M\}$. Since the characteristics of the candidate common ISS Lyapunov function (or the matrix $\bar{P}$) and its decay rate along the uncertain trajectories (or $\bar{Q}_v$ and $\gamma_d$) are independent of $\sigma$, we find that $\bar{V}$ is a valid common ISS Lyapunov function (see Definition~\ref{Definition:ISS-Lyapunov-Common}). Thus, we conclude robust ISS for the two-layer interconnected MAS of this paper under an arbitrarily fast switching of the control layer topology.\vspace{0.065in}\hfill $\blacksquare$
\begin{remark}
	In the absence of perturbations, i.e., when $d(t) = \textbf{0}$, we find $\dot{\bar{V}} \leq -x^T \bar{Q}_v x \prec 0$ (asymptotic convergence of all state trajectories to the origin). Using Rayleigh-Ritz inequality, we reach to $\lambda_{min}(\bar{P}) \|x\|^2 \leq \bar{V} \leq \lambda_{max}(\bar{P}) \|x\|^2$ and $\dot{\bar{V}} \leq - \lambda_{min}(\bar{Q}_v) \|x\|^2$, which guarantee the robust exponential convergence of all state trajectories to the origin: $\|x(t)\| \leq \kappa_e \exp^{-\sigma_e t} \|x(0)\|$ where $\kappa_e = \sqrt{\frac{\lambda_{max}(\bar{P})}{\lambda_{min}(\bar{P})}}$ and $\sigma_e = \frac{\lambda_{min}(\bar{Q}_v)}{2\lambda_{max}(\bar{P})}$ \cite{Khalil-Prentice-2003}.\vspace{0.065in}\hfill $\blacktriangleleft$
\end{remark}

\begin{remark}
	As an alternative to the proposed Design Procedure~\ref{DesignProcedure}, an interested reader can restrict the design to $R_i = r I_{n_u}$ with $r \in \mathbb{R}_{>0}$ for all agents, or to a set of $R_i$ and $\mathcal{H}_{c \sigma}$ that satisfy $\frac{\bar{R} \bar{E}_{c\sigma} + (\bar{R} \bar{E}_{c\sigma})^T}{2} \succcurlyeq \textbf{0}$ for all $\sigma$, in order to remove $2 \bar{K}^T \bar{R} \bar{E}_{c\sigma} \bar{K}$ from the validation matrix $\bar{Q}_{v \sigma}$  in~\eqref{eq:DesignCondition}. Consequently, that high-dimension validation matrix will turn into a set of $N$ low-dimension conditions $Q_{vi} = Q_i + K_i^T R_i K_i - \frac{1}{a_f} P_i B_{f_i} B_{f_i}^T P_i \succ \textbf{0}$ (one for each control node), and a $\sigma$-independent $\bar{Q}_v = \text{diag}\{Q_{vi}\}$ will directly appear (instead of $\bar{Q}_{v \sigma}$) in the proof of Theorem~\ref{Theorem:Main}.\hfill $\blacktriangleleft$
\end{remark}


\section{Simulation Verification}\label{Section:Simulation}
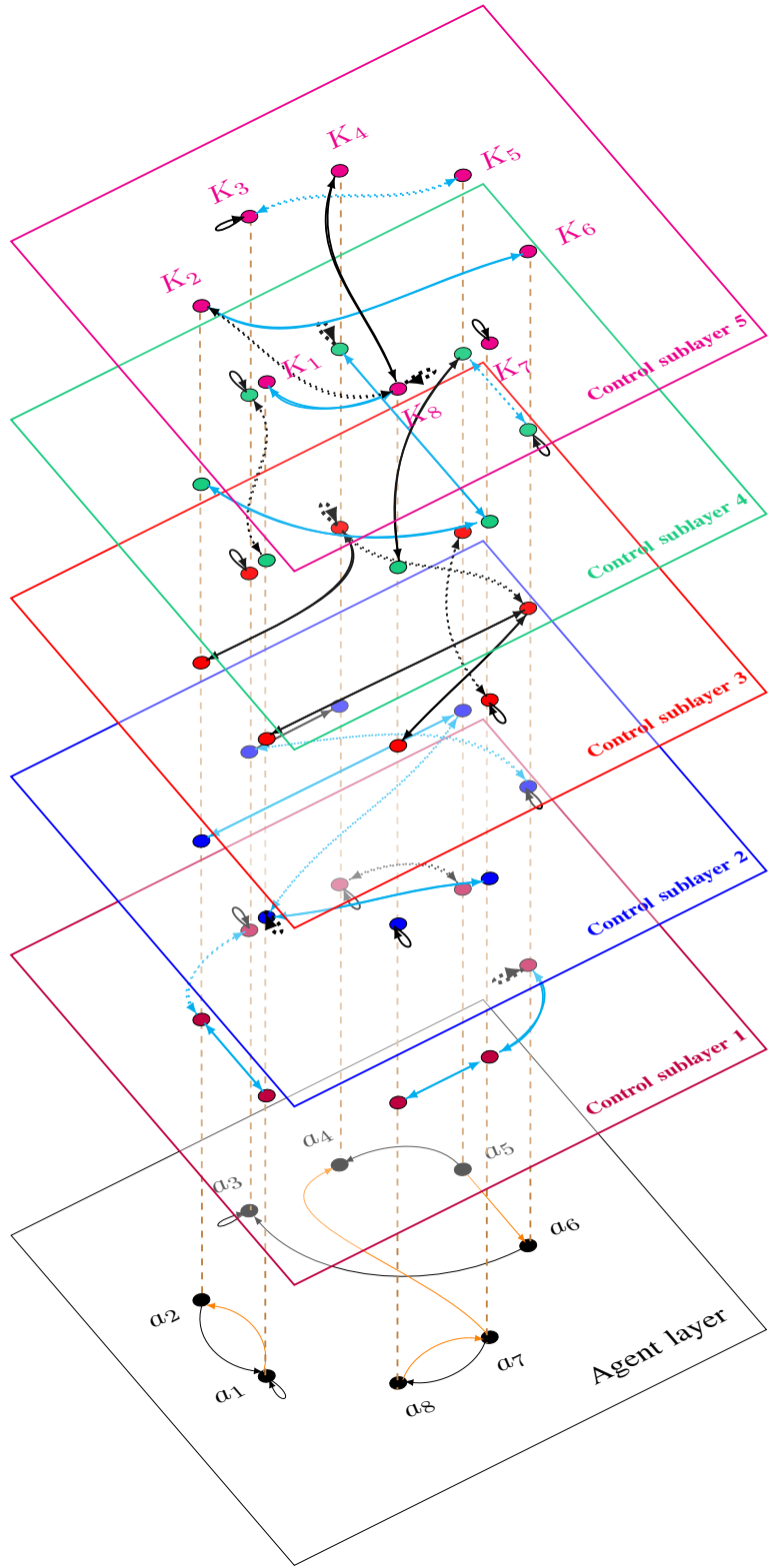
\begin{figure}[b!]
	\centering
	\resizebox{0.58\linewidth}{!}{
		\begin{tikzpicture}[scale=1.1,every node/.style={minimum size=1cm},on grid]
		\begin{scope}[
		yshift=-120,
		every node/.append style={yslant=\yslant,xslant=\xslant},
		yslant=\yslant,xslant=\xslant
		] 
		\draw[black, solid, thin] (0,0) rectangle (6.5,6.5); 
		\draw[fill=black]  
		(1.3,2.8) circle (0.1)     
		(1.3,4.3) circle (0.1)     
		(2.5,5.2) circle (0.1)     
		(3.75,5.21) circle (0.1) 
		(4.9,4.3) circle (0.1)     
		(4.9,2.8) circle (0.1)     
		(3.77,1.8) circle (0.1)   
		(2.51,1.8) circle (0.1);  
		
		\draw[-latex,thin,black] 
		(1.3,2.75) to[loop below] (1.3,2.75); 
		
		\draw[-latex,thin,orange] 
		(1.3,2.88) to[out = 45,in = -45] (1.3,4.2); 
		
		\draw[-latex,thin,black]
		(1.27,4.23) to[out = 225,in = 135] (1.3,2.88); 
		
		\draw[-latex,thin,black] 
		(2.45,5.2) to[loop left] (2.5,5.35); 
		
		\draw[-latex,thin,black]
		(4.8,2.8) to[out = 180,in = -90] (2.5,5.1); 
		
		\draw[-latex,thin,black] 
		(4.9,4.36) to[out = 90,in = 0] (3.81,5.21); 
		
		\draw[-latex,thin,orange] 
		(3.77,1.87) to[out = 90,in = 180] (3.67,5.21); 
		
		\draw[-latex,thin,orange] 
		(4.9,4.2) to[out = -90,in = 90] (4.9,2.85); 
		
		\draw[-latex,thin,orange] 
		(2.58,1.81) to[out = 45,in = 135] (3.7,1.81); 
				
		\draw[-latex,thin,black] 
		(3.7,1.78) to[out = -135,in = -45] (2.58,1.78); 
	
		\fill[black]
		(4.25,0.5) node[right, scale=1.1] {\text{Agent layer}}
		(1.3,2.8) node[left,scale=1.1]{$a_1$} 
		(1.3,4.3) node[left,scale=1.1]{$a_2$} 
		(2.5,5.2) node[above,scale=1.1]{$a_3$} 
		(3.75,5.21) node[above,scale=1.1]{$a_4$} 
		(4.9,4.3) node[right,scale=1.1]{$a_5$} 
		(4.9,2.8) node[right,scale=1.1]{$a_6$} 
		(3.77,1.8) node[below,scale=1.1]{$a_7$} 
		(2.51,1.8) node[below,scale=1.1]{$a_8$}; 
		\end{scope}
		
		\draw[thick, dashed, brown](-0.4, 12.11) to (-0.4,-1.6); 
		\draw[thick, dashed, brown](-1.3, 13.05) to (-1.27,-0.54); 
		\draw[thick, dashed, brown](-0.6, 14.35) to (-0.6,0.68); 
		\draw[thick, dashed, brown](0.64, 14.85) to (0.64,1.425); 
		\draw[thick, dashed, brown](2.32, 14.85) to (2.32,1.25); 
		\draw[thick, dashed, brown](3.25, 13.765) to (3.25,0.2); 
		\draw[thick, dashed, brown](2.65, 12.47) to (2.65,-1.15); 
		\draw[thick, dashed, brown](1.42, 11.85) to (1.42,-1.8); 
		%
		
		\begin{scope}[
		yshift= -10,
		every node/.append style={yslant=\yslant,xslant=\xslant},
		yslant=\yslant,xslant=\xslant
		]
		\fill[white,fill opacity=.35] (0,0) rectangle (6.5,6.5); 
		\draw[purple, solid, thick] (0,0) rectangle (6.5,6.5); 
		\draw[fill=purple]  
		(1.3,2.8) circle (0.1) 
		(1.3,4.3) circle (0.1) 
		(2.5,5.2) circle (0.1) 
		(3.75,5.21) circle (0.1) 
		(4.9,4.3) circle (0.1) 
		(4.9,2.8) circle (0.1) 
		(3.77,1.8) circle (0.1) 
		(2.51,1.8) circle (0.1); 
		
		\draw[-latex,solid,thick,cyan] 
		(1.3,2.88) to[out = 90,in = -90] (1.30,4.23); 
		\draw[-latex,solid,thick,cyan] 
		(1.30,4.23) to[out = -90,in = 90] (1.3,2.88); 

		\draw[-latex,dotted,thick,cyan] 
		(1.3,4.4) to[out = 90,in = 180] (2.45,5.21); 
		\draw[-latex,dotted,thick,cyan] 
		(2.45,5.21) to[out = 180,in = 87] (1.3,4.4); 

		\draw[-latex,solid,thick,black]
		(2.52,5.23) to[loop above] (2.5,5.135);          
		
		\draw[-latex,solid,thick,black]
		(3.75,5.13) to[loop below] (3.75,5.13); 
		\draw[-latex,dotted,thick,black] 
		(3.88,5.21) to[out = 0,in = 90] (4.9,4.418); 
		\draw[-latex,dotted,thick,black] 
		(4.9,4.418) to[out = 90,in = 0] (3.88,5.21); 
		
		\draw[-latex,dotted,ultra thick,black]
		(4.82,2.78) to[loop left] (4.82,2.78); 
		\draw[-latex,solid,thick,cyan] 
		(4.9,2.7) to[out = -90,in = 0] (3.9,1.8); 
		\draw[-latex,solid,thick,cyan] 
		(3.9,1.8) to[out = 0,in = -92] (4.9,2.7); 

		\draw[-latex,solid,thick,cyan] 
		(3.6,1.8) to[out = 1800,in = 0] (2.61,1.8); 
		\draw[-latex,solid,thick,cyan] 
		(2.61,1.8) to[out = 0,in = 180] (3.65,1.8); 

		\fill[purple]
		(4.16,0.35) node[right, scale=0.75] { \textbf{Control sublayer 1}}
		;
		\end{scope}
%
		\begin{scope}[
		yshift=60,
		every node/.append style={yslant=\yslant,xslant=\xslant},
		yslant=\yslant,xslant=\xslant
		]
		\fill[white,fill opacity=.35] (0,0) rectangle (6.5,6.5); 
		\draw[blue, solid, thick] (0,0) rectangle (6.5,6.5); 
		\draw[fill=blue]  
		(1.3,2.8) circle (0.1)   
		(1.3,4.3) circle (0.1)   
		(2.5,5.2) circle (0.1)   
		(3.75,5.21) circle (0.1) 
		(4.9,4.3) circle (0.1)   
		(4.9,2.8) circle (0.1)   
		(3.77,1.8) circle (0.1)  
		(2.51,1.8) circle (0.1); 
		
		\draw[-latex, dotted, ultra thick, black] 
		(1.325,2.89) to[loop below] (1.325,2.89); 
		\draw[-latex, solid, thick, black]
		(4.859,2.8) to[loop below] (4.79,2.8);      
		\draw[-latex, solid, thick, black] 
		(2.45,1.8) to[loop below] (2.45,1.8);    
						
		\draw[-latex,dotted,thick,cyan] 
		(1.375,2.85) to[out = 31.5,in = -148] (4.8,4.25); 
		\draw[-latex,dotted,thick,cyan] 
		(4.8,4.25) to[out = -148,in = 31.5] (1.375,2.85); 
		\draw[-latex,solid,thick,cyan] 
		(1.375,2.75) to[out = -27.5,in = 151] (3.687,1.8); 
		\draw[-latex,solid,thick,cyan] 
		(3.687,1.8) to[out = 151,in = -27.5] (1.375,2.75); 
		
		\draw[-latex,solid,thick,cyan] 
		(1.39,4.31) to[out = 0,in = -180] (4.78,4.3);  
		\draw[-latex,solid,thick,cyan] 
		(4.78,4.3) to[out = -180,in = 0] (1.39,4.31);  
		
		\draw[-latex,solid,thick,black] 
		(2.58,5.2) to[out = 0,in = 180] (3.657,5.21);      
		\draw[-latex,solid,thick,black] 
		(3.657,5.21) to[out = 180,in = 0] (2.6,5.2);       
		\draw[-latex,dotted,thick,cyan] 
		(2.58,5.2) to[out = -25,in = 90] (4.93,2.89);      
		\draw[-latex,dotted,thick,cyan] 
		(4.93,2.89) to[out = 90,in = -25] (2.58,5.2);       
		
		\fill[blue]
		(4.18,0.35) node[right, scale=0.75] { \textbf{Control sublayer 2}}
		;
		\end{scope}

		\begin{scope}[
		yshift=130,
		every node/.append style={yslant=\yslant,xslant=\xslant},
		yslant=\yslant,xslant=\xslant
		]
		\fill[white,fill opacity=.35] (0,0) rectangle (6.5,6.5); 
		\draw[red, solid, thick] (0,0) rectangle (6.5,6.5); 
		\draw[fill=red]
		(1.3,2.8) circle (0.1) 
		(1.3,4.3) circle (0.1) 
		(2.5,5.2) circle (0.1) 
		(3.75,5.21) circle (0.1) 
		(4.9,4.3) circle (0.1) 
		(4.9,2.8) circle (0.1) 
		(3.77,1.8) circle (0.1) 
		(2.51,1.8) circle (0.1); 
		
		\draw[-latex,solid,thick,black] 
		(2.5,5.26) to[loop above] (2.5,5.2); 
		
		\draw[-latex,dotted,ultra thick,black]
		(3.75,5.26) to[loop above] (3.75,5.21); 

		\draw[-latex,solid,thick,black] 
		(3.74,1.77) to[loop below] (3.77,1.8); 
		
		\draw[-latex,solid,thick,black] 
		(1.367,2.835) to[out = 0,in = 180] (4.855,2.8); 
		\draw[-latex,solid,thick,black] 
		(4.855,2.8) to[out = 180,in = 0] (1.367,2.835); 

		 \draw[-latex,thick,black] 
		 (1.367,4.3) to[out = 0,in = -96] (3.75,5.15); 
		 \draw[-latex,thick,black] 
		 (3.75,5.15) to[out = -90,in = 0] (1.367,4.3); 
		
		\draw[-latex,dotted,thick,black]
		(3.75,5.15) to[out = -90,in = 90] (4.89,2.88); 
		\draw[-latex,dotted,thick,black]
		(4.9,2.88) to[out = 90,in = -88] (3.745,5.15); 
		
		\draw[-latex,dotted,thick, black]
		(4.825,4.26) to[out = -135,in = 90] (3.77,1.87); 
		\draw[-latex,dotted,thick, black]
		(3.77,1.87) to[out = 90,in = -135] (4.825,4.26); 

		\draw[-latex,solid,thick,black] 
		(4.84,2.73) to[out = -155,in = 23.5] (2.58,1.84); 
		\draw[-latex,solid,thick,black] 
		(2.58,1.84) to[out = 23.5,in = -155] (4.84,2.73); 

		\fill[red]
		(4.18,0.35) node[right, scale=0.75] { \textbf{Control sublayer 3}}
		;
		\end{scope}
		
		\begin{scope}[
		yshift= 200,
		every node/.append style={yslant=\yslant,xslant=\xslant},
		yslant=\yslant,xslant=\xslant
		]
		\fill[white,fill opacity=.1] (0,0) rectangle (6.5,6.5); 
		\draw[darkgreen, solid, thick] (0,0) rectangle (6.5,6.5); 
		\draw[fill=darkgreen]
		(1.3,2.8) circle (0.1) 
		(1.3,4.3) circle (0.1) 
		(2.5,5.2) circle (0.1) 
		(3.75,5.21) circle (0.1) 
		(4.9,4.3) circle (0.1) 
		(4.9,2.8) circle (0.1) 
		(3.77,1.8) circle (0.1) 
		(2.51,1.8) circle (0.1); 
		
		\draw[-latex,solid,thick,black] 
		(2.5,5.26) to[loop above] (2.5,5.2);           
		
		\draw[-latex,dotted,ultra thick,black]
		(3.75,5.31) to[loop above] (3.75,5.21);     

		\draw[-latex,solid,thick,black]
		(4.9,2.7) to[loop below] (4.9,2.7);         
		
		\draw[-latex,dotted,thick,black] 
		(1.3,2.92) to[out = 90,in = -95] (2.5,5.12);  
		\draw[-latex,dotted,thick,black] 
		(2.5,5.12) to[out = -90,in = 85] (1.3,2.92);  

		\draw[-latex,solid,thick,cyan] 
		(1.35,4.25) to[out = -75,in = 165] (3.64,1.87); 
		\draw[-latex,solid,thick,cyan] 
		(3.64,1.87) to[out = 165,in = -75] (1.35,4.25);   
		
		\draw[-latex,thick,cyan]
		(3.75,5.14) to[out = -90,in = 90] (3.76,1.865);  
		\draw[-latex,thick,cyan]
		(3.76,1.865) to[out = 90,in = -90] (3.75,5.14);  
		
		\draw[-latex,dotted,thick,cyan]
		(4.92,4.17) to[out = -90,in = 90] (4.93,2.90); 
		\draw[-latex,dotted,thick,cyan]
		(4.93,2.90) to[out = 90,in = -90] (4.92,4.17);   

		\draw[-latex,solid,thick,black] 
		(4.84,4.27) to[out = 205,in = 65] (2.57,1.88);    
		\draw[-latex,solid,thick,black] 
		(2.57,1.88) to[out = 65,in = 205] (4.84,4.27);    
		
		\fill[darkgreen]
		(4.18,0.35) node[right, scale=0.75] { \textbf{Control sublayer 4}}
		;
		\end{scope}
		
		\begin{scope}[
		yshift= 270,
		every node/.append style={yslant=\yslant,xslant=\xslant},
		yslant=\yslant,xslant=\xslant
		]
		\fill[white,fill opacity=.1] (0,0) rectangle (6.5,6.5); 
		\draw[magenta, solid, thick] (0,0) rectangle (6.5,6.5); 
		\draw[fill=magenta]
		(1.3,2.8) circle (0.1)     
		(1.3,4.3) circle (0.1)     
		(2.5,5.2) circle (0.1)     
		(3.75,5.21) circle (0.1) 
		(4.9,4.3) circle (0.1)     
		(4.9,2.8) circle (0.1)     
		(3.77,1.8) circle (0.1)   
		(2.51,1.8) circle (0.1);  
		
		\draw[-latex,solid,thick,black] 
		(2.45,5.236) to[loop left] (2.5,5.2);               

		\draw[-latex,solid,thick,black] 
		(3.79,1.842) to[loop above] (3.79,1.842);           
		
		\draw[-latex,dotted, ultra thick,black] 
		(2.625,1.83) to[loop right] (2.54,1.873);            
		
		\draw[-latex,solid,thick,cyan] 
		(1.28,2.738) to[out = -90,in = 180] (2.451,1.8);   
		\draw[-latex,solid,thick,cyan] 
		(2.451,1.8) to[out = 180,in = -90] (1.28,2.738);   

		\draw[-latex,solid,thick,cyan] 
		(1.35,4.25) to[out = -75,in = 165] (4.83,2.8);   
		\draw[-latex,solid,thick,cyan] 
		(4.83,2.8) to[out = 165,in = -72.5] (1.35,4.25);   

		\draw[-latex,dotted,thick,black] 
		(1.35,4.25) to[out = -75,in = 165] (2.44,1.8); 
		\draw[-latex,dotted,thick,black] 
		(2.44,1.8) to[out = 165,in = -75] (1.35,4.25); 
		
		 \draw[-latex,dotted,thick,cyan] 
		 (2.58,5.2) to[out = 0,in = 180] (4.83,4.3); 
		 \draw[-latex,dotted,thick,cyan] 
		 (4.83,4.3) to[out = 180,in = 0] (2.58,5.2); 

		\draw[-latex,thick,black]
		(3.66,5.17) to[out = -135,in = 75] (2.55,1.91); 
		\draw[-latex,thick,black]
		(2.55,1.91) to[out = 75,in = -135] (3.66,5.17); 
		
		\fill[magenta]
		(4.18,0.35) node[right, scale=0.75] { \textbf{Control sublayer 5}};
		\fill[magenta]
		(1.25,2.8) node[right,scale=1.1]{$K_1$}          
		(1.3,4.3) node[above,scale=1.1]{$K_{2}$}          
		(2.5,5.2) node[above,scale=1.1]{$K_{3}$}     
		(4.1,5.151) node[above,scale=1.1]{$K_{4}$} 
		(4.9,4.3) node[right,scale=1.1]{$K_{5}$}        
		(5.0,2.75) node[right,scale=1.1]{$K_{6}$}        
		(3.77,1.8) node[below,scale=1.1]{$K_{7}$}    
		(2.51,1.8) node[below,scale=1.1]{$K_{8}$};   
		\end{scope}
		
				
		\end{tikzpicture}
	    }    
	\caption{A two-layer interconnected MAS with five control sublayers, where $a_i$ denotes the agent and $K_i$ the stabilization gains associated to each node $i$ according to~\eqref{eq:virtualstabilziation}. $K_i$ is the same for each control node, to be found by following the dashed vertical lines. In simulation of  Fig.~\ref{Fig:Sublayer1UnderAttack}, only control sublayer 1 is active. In simulation of Fig.~\ref{Fig:StableClosedLoopMAS}, all control sublayers are frequently active, one at each time, according to subplot (a) in that figure. For the sake of visibility, \textit{the color of each control sublayer's frame is the same as its nodes}. The \textit{dotted horizontal arrows denote the compromised (blocked) communication links} by distributed DoS attacks.}
	\label{Fig:TwoLayerMAS}
\end{figure}

Now we demonstrate the feasibility of the proposed ideas for an interconnected MAS~\eqref{eq:AgentModel}, with an (unknown) interconnection topology $\mathcal{G}_a$ shown over the agent layer in Fig.~\ref{Fig:TwoLayerMAS}, where each black arrow represents an edge weight equal to~$1$ and orange arrow~$-1$. Together with the agent dynamics in Appendix~I, this builds an unstable (open-loop) agent layer with divergent trajectories.

We follow the steps of the proposed Design Procedure~\ref{DesignProcedure} and SMT-based Algorithms~1 and~2, and develop a control layer with $M=5$ sublayers as depicted in Fig.~\ref{Fig:TwoLayerMAS}. In particular, as a few conditions in the proposed cybersecurity Algorithms~\ref{alg:main} and~\ref{alg:smt}, we set $T = 3$, $r_{1 4} = r_{3 7} = r_{3 8} = 0$, and $\theta_2 = \theta_5 = 0$ (other parameters are relatively evident). Each edge with a cyan color represents a weight equal to 2, and black equal to 4. We obtain the robust distributed ISS gains $K_1 = [-5.2632,   -6.6137]$, $K_2 = [-4.8272,   -6.3715]$, $K_3 = [-1.8623,   -3.8913]$, $K_4 = [-5.0504,   -6.4468]$, $K_5 = [5.0014,    5.1293]$, $K_6 = [5.1795,    4.9353]$, $K_7 = [5.3290,    5.4146]$, and $K_8 = [3.8166,    3.7070]$ such that the validation condition~\eqref{eq:DesignCondition} is satisfied.

\begin{figure}[b!]
  \centering
  \includegraphics[width=0.90\linewidth]{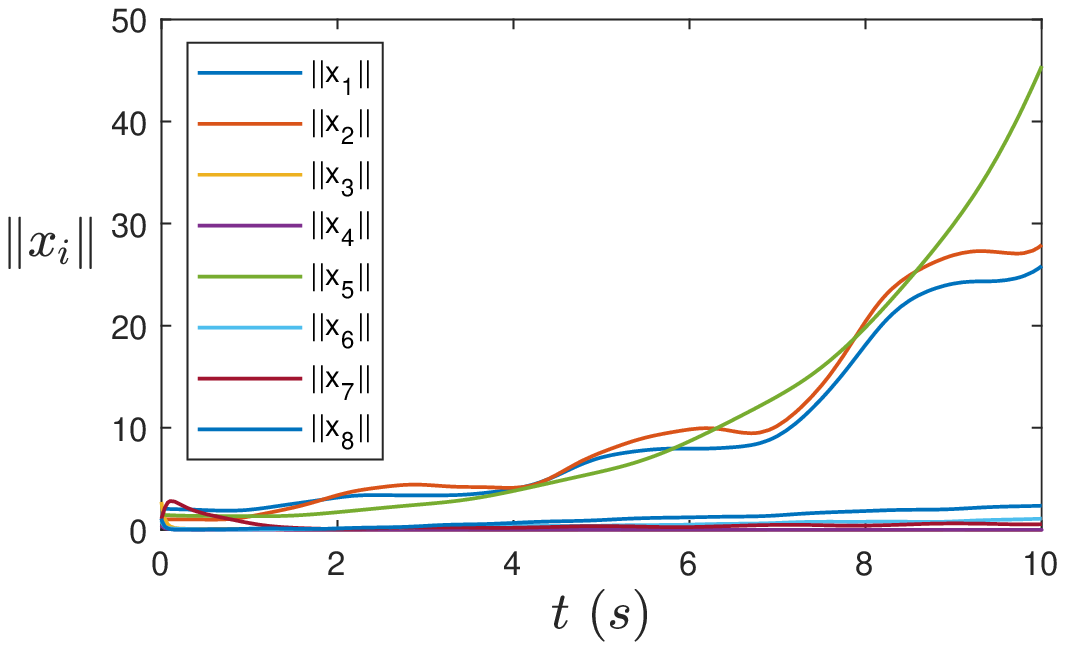}  
  \caption{Compromised system: Divergent trajectories show that a smart attacker, with limited resources, can easily compromise a fixed (predictable) control sublayer 1.\vspace{-0.0in}}
\label{Fig:Sublayer1UnderAttack}
\end{figure}

It is a common practice to assume that an attacker does not have unlimited resources \cite{jajodia2011moving}. Thus, we assume three communication links (including one selfloop) can be compromised per each sublayer (see dotted arrows over the control sublayers in Fig.~\ref{Fig:TwoLayerMAS}). Despite this assumption, the divergent trajectories in Fig.~\ref{Fig:Sublayer1UnderAttack} demonstrate that a smart attacker can easily compromise the underlying two-layer interconnected MAS if its control layer topology is fixed.

\begin{figure}[t!]
\centering
\begin{subfigure}{.979\textwidth}
  \centering
  \includegraphics[width=0.90\linewidth]{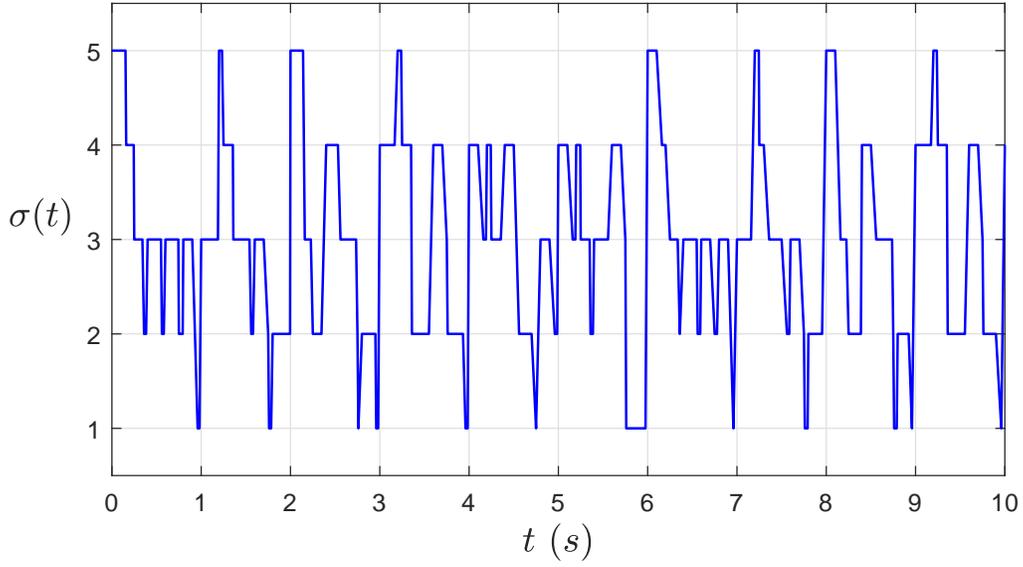}  
  \caption{The switching signal $\sigma$ indicates the number of an active control sublayer $1$ to $5$ for an arbitrarily fast switching.}
  \label{fig:sub-first}
\end{subfigure}
\begin{subfigure}{.95\textwidth}
  \centering
  \includegraphics[width=0.950\linewidth]{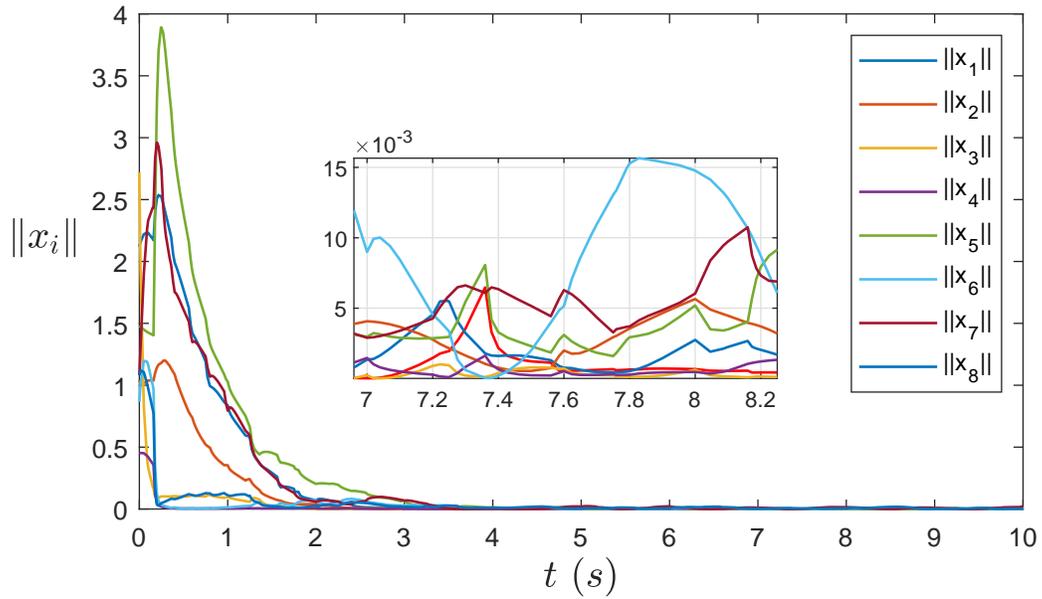}  
  \caption{Robust ISS behavior under the distributed DoS, modeling uncertainties, nonvanishing perturbations, and very fast switching.}
  \label{fig:sub-second}
\end{subfigure}
\caption{Proactive cyber defense: Robust ISS for the two-layer interconnected MAS in Fig.~\ref{Fig:TwoLayerMAS} under a (time-varying) \textcolor{black}{persistent}, distributed DoS attack, as explained in Section~\ref{Section:Simulation}.\vspace{-0.0in}}
\label{Fig:StableClosedLoopMAS}
\end{figure}

To examine the power of the proposed proactive cyber defense strategy, we consider an attack scenario where the attacker (\textcolor{black}{persistently}) attacks on (the dotted edges of) the control sublayer 5 over $t \in [0,2)$, 4 over $t \in [2,4)$, 3 over $t \in [4,6)$, 2 over $t \in [6,8)$, and 1 over $t \in [8,10]$, in addition to the permanently compromised (potential) communication capabilities between the control nodes $1$ and $4$, $3$ and $7$, and $3$ and $8$. Note that the two-layer interconnected MAS starts under a DoS attack, and persistently remains under attack. This is different from the existing switching-based studies \cite{Rezaei-CDC-2021} and \cite{FengTesiDePersis-CDC-2017}, where the attack frequently goes off and, in average, the system has (enough) time to recover after each DoS. As shown in Fig.~\ref{Fig:StableClosedLoopMAS}, via an arbitrarily fast switching of the control sublayers, the proposed distributed protocol ensures ISS for the two-layer interconnected MAS in the presence of various abnormalities.

\clearpage
\newpage 

All individual subsystems should be asymptotically stable in order to guarantee the asymptotic stability of an arbitrarily fast switched system (subsection 2.1.1, \cite{Liberzon-Springer-2003}). In the ISS simulation results of this section, the two-layer interconnected MAS may temporarily face an attack on a few of its communication links, which violates such a condition. This could be problematic pointing to the fact that the underlying agent layer is unstable. However, due to the ``non-overlapping paths" and ``centrality distribution" conditions in the developments of Subsection~\ref{Subsection:MDT}, a fast mutation among the control sublayers ensures that the final moving target defense strategy will end in a two-layer interconnected MAS that (with an abuse of the words) is ``more (input-to-state) stable than unstable." Indeed, over a time interval, the agent layer dynamics see a (time varying) control layer topology for which the conditions of Section~\ref{Section:Preliminaries} are satisfied by the union of the activated control sublayers. Further computer science-oriented investigation on this subject is left for the future. However, without any technical modifications, we need to mention that an increase in the number of nodes (in simulation) will increase the power of the proposed moving target defense strategy to handle the (unmeasured) cyber attacks in a proactive manner, because it will increase the number of possible control sublayer structures that would satisfy the conditions of Subsection~\ref{Subsection:MDT}.


\section{Summary}\label{Section:Summary}
We systematically study the robust input-to-state stability and proactive security in an interconnected multiagent system (subject to multiple cyber and physical abnormalities), based on a synergistic combination of various concepts from the literature of controls, graph theory, and computer science. In particular, we design a set of cybersecurity-aware, robust control sublayers based on a mixed optimal control, graph, and satisfiability modulo theory formulation. Then, relying on the arbitrarily fast switching capability of the proposed distributed stabilization protocol and the designed cybersecurity-aware control sublayers, we offer a moving target defense strategy, and enhance the cybersecurity aspect of the two-layer interconnected multiagent system in a proactive manner. The proposed systematic framework may pave the way for an effective and comprehensive study of the cyber-physical multiagent systems from both control and cybersecurity viewpoints \cite{ChongSandbergTeixeira-ECC-2019}, with an application to power systems (with their inherently interconnected dynamics) \cite{ZhouShahidehpourPaasoBahramiradAlabdulwahabAbusorrah-CST-2020}.

\section*{Appendix I}
In Fig.~\ref{Fig:TwoLayerMAS}, according to model~\eqref{eq:AgentModel}, the (unstable) nominal part of interconnected MAS is characterized by the following matrices for $l \in \{1,2,3,4\}$ and $m \in \{5,6,7,8\}$:
\begin{equation*}
    \begin{array}{rl}
         & A_l  = \begin{bmatrix} 0 & 1 \\ -1 & 0.25 \end{bmatrix} ~~~~~ B_{u_l} = \begin{bmatrix} 0 \\ 1 \end{bmatrix} \\
         & A_m = \begin{bmatrix} 0 & 1 \\ 0.25 & -1 \end{bmatrix} ~~~ B_{u_m} = \begin{bmatrix} 0.25 \\ -1 \end{bmatrix}
     \end{array}   
\end{equation*}
\normalsize
and the modeling uncertainty and perturbation matrices are:
\begin{equation*}
    \begin{array}{rl}
		 B_{f_1} & = B_{u_1} ~~~~~~~~~ B_{d_1} = B_{u_1}  ~~~\text{(matched scenario)}\\
          B_{f_2} & = \begin{bmatrix} 0.5 \\ -1\end{bmatrix} ~~~~~~~ B_{d_2} = \begin{bmatrix} 0.25 \\ -0.75 \end{bmatrix}\\
          B_{f_3} & = \begin{bmatrix} 0.25 \\ -0.75\end{bmatrix} ~~~ B_{d_3} = \begin{bmatrix} 0.5 \\ 1 \end{bmatrix}\\
          B_{f_4} & = B_{u_4} ~~~~~~~~~ B_{d_4} = B_{u_4}  ~~~\text{(matched scenario)} \\
          B_{f_5} & = \begin{bmatrix} -0.5 \\ 0.5\end{bmatrix} ~~~~~ B_{d_5} = B_{u_5}\\
          B_{f_6} & = \begin{bmatrix} 0\\ 1\end{bmatrix} ~~~~~~~~~ B_{d_6} = B_{f_6}\\
          B_{f_7} & = \begin{bmatrix} 0 \\ -1\end{bmatrix} ~~~~~~ B_{d_7} = \begin{bmatrix} 0.5 \\ 0.5 \end{bmatrix}\\
          B_{f_8} & = B_{u_8} ~~~~~~~~~ B_{d_8} = \begin{bmatrix} 0\\ 1 \end{bmatrix}.
    \end{array}
\end{equation*} 
\normalsize

Also, the (unknown) nonlinearities and interconnection matrices are $f_1(z_1) = 0.5 \tanh(z_1)$, $f_2 (z_2) = -0.4 \sin(z_2)$, $f_3 (z_3,t) = 0.5 \sin(t) \tanh(z_3)$, $f_4 (z_4) = -0.4 \tanh(z_4) $, $f_5 (z_5) = -0.5 \sin(z_5)$, $f_6 (z_6,t) = 0.4 \sin(t) \sin(z_6)$, $f_7(z_7) = 0.5 z_7$, and $f_8 (z_8)= 0.4 \tanh(z_8)$, as well as $C_l = [0, 1]$ for $l \in \{1,2,3,4\}$, and $C_m = [-1,0]$ for $m \in \{5,6,7,8\}$. The nonmeasurable, nonvanishing, external perturbations are $d_p(t) = \frac{1}{3} \sin(\pi t)$, and $d_q(t) = \frac{1}{3} \sin(2\pi t)$ for $p \in \{1,3,5,7\}$ and $q \in \{2,4,6,8\}$.


\end{document}